# Electron transport properties of a narrow-bandgap semiconductor $Bi_2O_2Te$ nanosheet

Xiaobo Li[1,2], Haitian Su[1], and H. Q. Xu[1,3,*]

[1]*Beijing Key Laboratory of Quantum Devices, Key Laboratory for the Physics and Chemistry of Nanodevices, and School of Electronics, Peking University, Beijing 100871, China*

[2]*Academy for Advanced Interdisciplinary Studies, Peking University, Beijing 100871, China*

[3]*Beijing Academy of Quantum Information Sciences, Beijing 100193, China*

[*]Corresponding author; email: hqxu@pku.edu.cn

(Dated: May 26, 2022)

**ABSTRACT**

A thin, narrow-bandgap semiconductor $Bi_2O_2Te$ nanosheet is obtained via mechanical exfoliation and a Hall-bar device is fabricated from it on a heavily doped $Si/SiO_2$ substrate and studied at low temperatures. Gate transfer characteristic measurements show that the transport carriers in the nanosheet are of *n*-type. The carrier density, mobility, and mean free path in the nanosheet are determined by measurements of the Hall resistance and the longitudinal resistance of the Hall-bar device and it is found that the electron transport in the nanosheet is in a quasi-two-dimensional (2D), strongly disordered regime. Magnetotransport measurements for the device at magnetic fields applied perpendicular to the nanosheet plane show dominantly weak antilocalization (WAL) characteristics at low fields and a linear magnetoresistance (LMR) behavior at large fields. We attribute the WAL characteristics to strong spin-orbit interaction (SOI) and the LMR to the classical origin of strong disorder in the nanosheet. Low-field magnetoconductivity measurements are also performed and are analyzed based on the multi-channel Hikami-Larkin-Nagaoka theory with the LMR correction being taken into account. The phase coherence length, spin relaxation length, effective 2D conduction channel number and coefficient in the linear term due to the LMR in the nanosheet are extracted. It is found that the spin relaxation length in the $Bi_2O_2Te$ nanosheet is several times smaller than it in its counterpart $Bi_2O_2Se$ nanosheet and thus an ultra-strong SOI is present in the $Bi_2O_2Te$ nanosheet. Our results reported in this study would greatly encourage further studies and applications of this emerging narrow-bandgap semiconductor 2D material.

Since discovery of graphene,[1] two-dimensional (2D) materials have drawn extensive attentions due to their rich, novel physical properties and potential applications in electronics, optoelectronics and quantum electronics.[1-6] Recently, 2D semiconductor $Bi_2O_2Se$ has caught an increasing research interest. This 2D material can be obtained via chemical vapor deposition (CVD) and has shown to exhibit several superior properties, such as ultrahigh electron mobility, excellent optical responses, and ambient stability.[7-10] It is also experimentally demonstrated that thin $Bi_2O_2Se$ layers possess strong spin-orbit interaction (SOI)[11,12] and could thus be employed to construct a planar superconducting Josephson junction[13] as well as a planar topological quantum device.[14,15] Semiconductor $Bi_2O_2Te$ is another bismuth oxychalcogenide and has a similar crystal structure as $Bi_2O_2Se$,[16-19] But, as a most interesting aspect, $Bi_2O_2Te$ has a relatively small bandgap (~0.23 eV)[16] and is therefore expected to exhibit an even stronger SOI and to be an even more favorable material for constructing a planar topological superconducting junction device.[20-22] So far, 2D $Bi_2O_2Te$ has rarely been investigated by transport measurements and the carrier transport characteristics in this material remain largely unknown.

Here, we report on an experimental study of the transport properties of a thin $Bi_2O_2Te$ nanosheet. The thin nanosheet is obtained by mechanical exfoliation and is casted into a Hall bar on a $SiO_2$/Si substrate. Gate transfer characteristic measurements show that the transport carriers in the nanosheet are of *n*-type. The Hall resistance and the longitudinal resistance are measured at low temperatures with magnetic fields applied perpendicular to the nanosheet plane. The sheet carrier density and the mobility in the nanosheet at different gate voltages are extracted to be $\sim 2\times10^{14}$ cm$^{-2}$ and ~36 cm$^2$/V·s, respectively, and the carrier transport in the nanosheet is in a quasi-2D, strong disorder regime. The measured longitudinal resistance exhibits dominantly weak antilocalization (WAL) characteristics in the vicinity of zero magnetic field and a linear magnetoresistance (LMR) behavior at large fields. The observation of the WAL characteristics demonstrates that the $Bi_2O_2Te$ nanosheet possesses a strong SOI. Detailed low-field magnetotransport measurements are performed at different gate voltages and low temperatures. The measurements are analyzed based on a multi-channel Hikami-Larkin-Nagaoka (HLN) formula[11,23-25] with taking the LMR correction into account[26,27] and the quantum transport characteristic parameters in the nanosheet are extracted. It is found that the spin relaxation length in the $Bi_2O_2Te$ nanosheet is several times smaller than it in its counterpart $Bi_2O_2Se$ nanoplate and thus an ultra-strong SOI is present in the $Bi_2O_2Te$ nanosheet.

Semiconductor $Bi_2O_2Te$ has a similar crystal structure as $Bi_2O_2Se$, but with relatively larger lattice parameters of a~3.98 Å and c~12.70 Å,[16,18] as depicted in the left panel of Fig. 1(a). Due to weak electrostatic interaction between the $Bi_2O_2$ layers and the Te layers, $Bi_2O_2Te$ nanosheets can be obtained via mechanical exfoliation. In this work, the $Bi_2O_2Te$ nanosheets are mechanically exfoliated from a commercially available $Bi_2O_2Te$ bulk. The exfoliated $Bi_2O_2Te$ nanosheets are transferred onto a 300-nm $SiO_2$ capped, highly *n*-doped Si substrate (to be employed as the back gate to the nanosheets). For the electron transport property study, a few thin $Bi_2O_2Te$ nanosheets are selected and are casted into Hall-bar devices.[12] In each Hall-bar device, two large contact electrodes (source and drain) and four narrow probe electrodes (voltage probes) are fabricated on a selected nanosheet by contact pattern definition via electron-beam lithography, deposition of Ti/Au (5 nm/100 nm in thickness) via electron-beam evaporation, and lift-off. Here, we note that to improve the metal contact to the nanosheet, the samples are briefly dipped in de-ionized water-diluted $(NH_4)_2S_x$ solution before the metal deposition. The right panel of Fig 1(a) shows a scanning electron microscope (SEM) image of a fabricated device studied in this work and the circuit setup for transport measurements. In the device, the nanosheet has a thickness of $d$~27 nm determined by atomic force microscopy measurements (see Supplementary Materials). The distance $L$ between two voltage probes along the transport $x$ direction is designed to be ~2.6 μm and the width of the Hall bar $W$ (i.e., the distance between two voltage probes along the transverse $y$ direction) is ~4.7 μm. The transport measurements are performed in a physical property measurement system equipped with a uniaxial magnet. In the measurements, both longitudinal voltage $V_{xx}$ and transverse voltage $V_{yx}$ are detected using a standard lock-in technique in which an ac current $I$ of 100 nA at a frequency of 17 Hz is supplied between the source and drain contacts. The longitudinal resistance $R_{xx}$ and the Hall resistance $R_{yx}$ are obtained as $R_{xx} = \frac{V_{xx}}{I}$ and $R_{yx} = \frac{V_{yx}}{I}$.

The fabricated device is first characterized by gate transfer characteristic measurements. Figure 1(b) shows the sheet resistance $R$ (defined as $R = R_{xx} \times \frac{W}{L}$) measured as a function of back-gate voltage $V_g$ for the device shown in Fig. 1(a) at $T = 2$ K and $B = 0$ T. It is found that $R$ increases monotonically with decreasing $V_g$, indicating that the transport carriers in the nanosheet are of *n*-type. However, $R$ is seen to increase only by ~12% as $V_g$ decreases from 30 V to −30 V. This relatively weak $V_g$ dependence reflects the fact that that the nanosheet is heavily doped. Previously, an *n*-type $Bi_2O_2Te$ pellet was synthesized.[16] But, the physical origin of the *n*-type character of carriers in $Bi_2O_2Te$ has not yet been experimentally identified. A theoretical study predicted that although it is natural to grow *n*-type $Bi_2O_2Se$, as-grown $Bi_2O_2Te$ without doping should behave as an intrinsic semiconductor.[28]

Figure 1(c) shows the measured longitudinal resistance $R_{xx}$ and Hall resistance $R_{yx}$ of our $Bi_2O_2Te$ Hall-bar device at perpendicularly applied magnetic fields $B$ and at $T = 2$ K and $V_g = 0$ V. It is seen that $R_{xx}$ displays a well-defined dip at low fields, i.e., a typical WAL characteristc,[11, 25] revealing the presence of a strong SOI in the $Bi_2O_2Te$ nanosheet. Here we note that similar WAL characteristics are also observed in other Hall-bar devices fabricated from thicker $Bi_2O_2Te$ nanosheets (see Supplementary Materials). At large magnetic fields, $R_{xx}$ increases linearly with increasing $|B|$, see the plot for $R_{xx}$ at $1\,T < |B| < 2\,T$ in Fig. 1(c). We fit $R_{xx}$ to the formula $R_{xx}(B) = R_{xx0} + |\lambda| \cdot |B|$ at $1\,T < |B| < 2\,T$ and obtain $|\lambda| = 7\,\Omega \cdot T^{-1}$ and $R_{xx0} = 530\,\Omega$. Such a linear magnetoresistance (LMR) has been observed in previous studies[29-32] and has been explained in terms of a quantum or a classical model. In the quantum model, the LMR arises when all the carriers coalesce into the lowest Landau level in a material with a linear dispersion and zero bandgap.[30,33,34] As for the classical model, the LMR is considered to be originated from distortions in current paths caused by spatial fluctuations in the carrier mobility and is closely related to the disorder in a material.[35, 36] The observed LMR in our work could be attributed to the classical origin, considering the band structure of $Bi_2O_2Te$ and the fact that defects have been found to be a cause for the LMR in a bismuth oxychalcogenide.[37] The Hall resistance $R_{yx}$ is found to decrease linearly as $B$ increases, which indicates again that transport carriers in the $Bi_2O_2Te$ nanosheet are of electrons.

Similar measurements of the longitudinal resistance $R_{xx}$ and the Hall resistance $R_{yx}$ have also been carried out for the $Bi_2O_2Te$ Hall-bar device at different back-gate voltages $V_g$ and the similar characteristic results as in Fig. 1(c) are observed. Figure 1(d) shows the sheet electron density $n$ and the Hall mobility $\mu$ in the $Bi_2O_2Te$ nanosheet extracted as a function of $V_g$ from these measurements. It is seen that as $V_g$ decreases from $30$ V to $-30$ V, the carrier density $n$ is changed from $1.93 \times 10^{14}\,cm^{-2}$ to $1.73 \times 10^{14}\,cm^{-2}$, but the mobility $\mu$ remains nearly unchanged at $\sim 36\,cm^2/V \cdot s$. These results give an electron mean free path in the nanosheet on the order of $L_e \sim 8$ nm. Such a small mobility and a small mean free path reflect that electrons suffer from strong scattering, which is consistent with our inference that a high concentration of defects is present in the nanosheet and the system is in a strong disorder regime. After converting the obtained sheet electron density to the bulk electron density in the nanosheet, the electron Fermi wavelength $\lambda_F$ is estimated to be $\sim 5$ nm. In view of the nanosheet thickness $d \sim 27$ nm, it is reasonable to assume that there exist more than one but a small number of 2D subbands in the nanosheet and the electron transport in the system is in a quasi-2D regime.

Figure 2(a) shows the measured low-field magnetoconductivity of the $Bi_2O_2Te$ Hall-bar device, $\Delta\sigma_{xx}(B) = \sigma_{xx}(B) - \sigma_{xx}(0)$, at $T = 2$ K and at different $V_g$ (black opened circles). Here, $\sigma_{xx}$ is obtained from $\sigma_{xx} = \frac{R_{xx}}{R_{xx}^2+R_{yx}^2} \times \frac{L}{W}$. The bottom data points are measured at $V_g = -30$ V and the data points measured at all other back-gate voltages are successively vertically offset for clarify. These measurements all show a distinct WAL peak at $B$=0 T. But the peak broadens as $V_g$ decreases. The WAL arises from quantum interference in the $Bi_2O_2Te$ nanosheet. The correction to the low-field magnetoconductivity due to quantum interference in a quasi-2D, disordered conductor can be described by the multi-channel Hikami-Larkin-Nagaoka (HLN) theory.[23] However, our measured magnetoconductivity may also include a classical correction originating from the LMR, which could well persist at low magnetic fields in the strongly disordered system.[29,35] Considering the fact that $R_{xx} \gg R_{yx}$ and $|\lambda| \cdot |B| \ll R_{xx0}$ in our nanosheet, to the lowest order, the classical correction gives a linear magnetic field term in the magnetoconductivity. Overall, the correction to the low-field magnetoconductivity of the device can be expressed as[26,27]

$$\Delta\sigma_{xx}(B) = -N \cdot \frac{e^2}{\pi h} \left[ \frac{1}{2}\Psi\left(\frac{B_\varphi}{B}+\frac{1}{2}\right) + \Psi\left(\frac{B_e}{B}+\frac{1}{2}\right) - \frac{3}{2}\Psi\left(\frac{(4/3)B_{so}+B_\varphi}{B}+\frac{1}{2}\right) - \frac{1}{2}\ln\left(\frac{B_\varphi}{B}\right) - \ln\left(\frac{B_e}{B}\right) + \frac{3}{2}\ln\left(\frac{(4/3)B_{so}+B_\varphi}{B}+\frac{1}{2}\right) \right] - \gamma B, \qquad (1)$$

where $\Psi$ is the digamma function, $B_i = \frac{\hbar}{4eL_i^2}$ (where i=$\varphi$, so, e) are the characteristic fields with $L_i$ being the characteristic lengths, $N$ is the number of effective 2D conduction channels, and $\gamma = \frac{L}{W} \cdot \frac{\lambda}{R_{xx0}^2}$ is the coefficient of the classical correction to the magnetoconductivity due to the LMR.

We take $L_\varphi$, $L_{so}$, $L_e$, $N$ and $\gamma$ as fitting parameters to fit the magnetoconductivity data measured for the device at different $V_g$ to Eq. (1). The yellow dashed lines in Fig. 2(a) show the results of the fits. It is seen that all the measured magnetoconductivity data are excellently fitted by Eq. (1). Note that we have also fitted the measured magnetoconductivity data obtained for the nanosheet device at $T$=2 K and $V_g$=0 V to Eq. (1) in a larger magnetic field range of $|B|$<2 T and found that the fit is also excellent (see Supplementary Materials). Note also that it is worthwhile to check whether our measured magnetoresistivity data points could be well fitted by the theory of the quantum interference correction to the resistivity in a three dimensional (3D) disordered system.[38,39] We have made such attempts and found that no satisfactory fits could ever be achieved for our measurements (see Supplementary Materials). Thus, the observed WAL characteristics in the $Bi_2O_2Te$ nanosheet are analyzed in terms of the

multi-2D channel quantum interreference theory. Figure 2(b) shows the extracted values of $L_\varphi$, $L_{so}$ and $L_e$ from the fits using Eq. (1). It is seen that $L_\varphi$ is weakly dependent on $V_g$, as it is changed from 66 nm to 60 nm only with decreasing $V_g$ from 30 V to −30 V. A smaller $L_\varphi$ at a lower $V_g$ reflects a stronger dephasing process, which can be attributed to a reduction in Coulomb screening and thus an enhancement in electron-electron interaction[40] at a lower electron density. In addition, $L_\varphi$ is found to be larger than the nanosheet thickness $d$, which guarantees the validity of our analyses of the measured magnetoconductivity data based on a multi-2D channel theory. The extracted $L_{so}$ is ~27 nm and is nearly $V_g$-independent. Since the presence of inversion symmetry in this material, the observed SOI is considered to be of the Rashba type and could be tuned by an electric field across the nanosheet. Here, due to the fact that the device is made in a single-gate structure, it is hard to achieve a sufficiently large tuning of the electric field in the nanosheet.[11,25] The extracted $L_e$ is ~8 nm and is also weakly dependent on $V_g$. This extracted value of $L_e$ is consistent with the results determined above from the carrier density and mobility measurements shown in Fig. 1(d).

Figure 2(c) shows the extracted $N$ and $\gamma$ from the fits shown in Fig. 2(a). It is found that $N$ increases from 2.9 to 4.0 as $V_g$ decreases from 30 V to −30 V. From our estimation, the number of 2D conduction channels in the nanosheet could be $\frac{d}{\lambda_F} \sim 5$. Our extracted number $N < \frac{d}{\lambda_F}$ indicates that the 2D channels in the nanosheet are partially coherently coupled with each other. As $V_g$ decreases, the coherent coupling between the channels weakens, since $L_\varphi$ reduces, and thus the number of effective 2D conduction channels $N$ increases.[41] The extracted $\gamma$ is ~0.43 $\frac{e^2}{h} \cdot T^{-1}$ and is independent of $V_g$. This value is comparable to the value of $\gamma \approx$ 0.36 $\frac{e^2}{h} \cdot T^{-1}$ that would be obtained from the above determined values of $|\lambda|$ and $R_{xx0}$ based on the linear fit to the $R_{xx}$-$B$ curve.

The magnetoconductivity $\Delta\sigma_{xx}$ of the thin $Bi_2O_2Te$ device is also measured at different temperatures $T$ and analyzed using the multi-channel HLN theory. Figures 3(a) and 3(d) show these measurements (black opened circles) for the device at fixed gate voltages $V_g$ = 30 V and $V_g$ = −30 V, respectively. The bottom data points in Figs. 3(a) and (d) are measured at $T =$ 2 K and the data points measured at all other temperatures are successively vertically offset for clarify. At each $V_g$, a WAL magnetoconductivity peak appears in all considered temperatures $T$. But it gets broadened as $T$ increases. Again, the measured magnetoconductivity data are fitted to Eq. (1) and the yellow dashed lines in Figs. 3(a) and 3(d) are the results of the fits. Figures 3(b) and 3(e) show the extracted $L_\varphi$, $L_{so}$ and $L_e$ from the fits. At $V_g =$ 30 V (−30 V), $L_\varphi$ is seen to decrease from 66 nm (60 nm) to 36 nm (33 nm) as $T$ increases from 2 K to 12 K. it is also seen that $L_\varphi$ in both cases follow a power-law temperature dependence, $L_\varphi \sim T^{-0.51}$, which indicates that the dephasing process in the nanosheet is dominantly caused by electron-electron scattering with small-energy transfers.[40] Strong deviations from the power-law temperature dependence at $T \leq 5$ K are likely due to that at $T \leq 5$ K, the electron temperature in the device is actually higher than the cryostat temperatures.[25] The extracted $L_{so}$ values at both $V_g = 30$ V and $V_g = -30$ V are rather weakly dependent on $T$. This is consistent with the fact that the SOI in the semiconductor nanosheet arises from the presence of an inversion asymmetry or an electrical field and both should be temperature independent at the low temperatures considered. The extracted $L_e$ also shows a rather weak temperature dependence, fully consistent with the fact that electron scattering is mainly due to impurities and defects whose configurations in the nanosheet would remain unchanged in this low temperature range. Figures 3(c) and 3(f) show the extracted $N$ and $\gamma$ from the fits shown in Figs. 3(a) and 3(d). At both $V_g = 30$ V and $V_g = -30$ V, $N$ increases with increasing $T$. This is due to the fact that with increasing $T$, $L_\varphi$ is deceased and thus coherent coupling between 2D channels in the nanosheet weakens. Parameter $\gamma$ is also found to be independent of $T$ at both $V_g = 30$ V and $V_g = -30$ V. This is again as expected, since the degree of disorder is essentially the same within this low temperature range.

In conclusion, a back-gated Hall-bar device has been fabricated from a mechanically exfoliated thin $Bi_2O_2Te$ nanosheet and the transport properties of the thin $Bi_2O_2Te$ nanosheet have been investigated at low temperatures. The gate transfer characteristic measurements show that the electrical carriers in the nanosheet are of *n*-type. The Hall resistance and the longitudinal resistance measurements are employed to extract the electron density and mobility in the thin nanosheet at different gate voltages, and it is found that the electron transport in the system is in a quasi-2D, diffusion regime. The measured longitudinal resistance exhibits WAL characteristics and an LMR behavior. The observation of the WAL demonstrates the presence of SOI in the $Bi_2O_2Te$ nanosheet. The magnetoconductivity measurements have also been carried out at different gate voltages and different temperatures. The measurements are analyzed based on a multi-channel HLN formula with the LMR being taken into account and the characteristic transport lengths in the nanosheet are extracted. The extracted dephasing length $L_\varphi$ is found to be both gate voltage and temperature dependent. In particular, $L_\varphi$ shows a power-law temperature dependent, revealing that dephasing in the nanosheet is dominantly due to electron-electron scattering with small energy transfers. The extracted spin relaxation length $L_{so}$ and mean free path $L_e$ show a weak dependence on gate voltage and temperature. Most importantly, a very short length of $L_{so} \sim 28$ nm is extracted in this work. Thus, the SOI in the $Bi_2O_2Te$ nanosheet is exceptionally strong. Our results presented here provide a solid basis for further studies as well as applications of the emerging thin, narrow-bandgap, semiconductor $Bi_2O_2Te$ materials.

**SUPPLEMENTARY MATERIAL**

The supplementary material file contains additional measurement data for the $Bi_2O_2Te$ nanosheet Hall-bar device studied in the main article, an evidence of failure in analyzing the magnetotransport measurements shown in the main article using the quantum interference theory developed for a three-dimensional disordered system, results of fit of the measured magnetoconductivity data to the Hikami-Larkin-Nagaoka (HLN) formula with the linear magnetoresistance (LMR) effect being taken into account in a larger field range of |B|<2 T, and magnetotransport measurement data for additional $Bi_2O_2Te$ nanosheet Hall-bar devices.

**ACKNOWLEDGEMENTS**

This work is supported by the Ministry of Science and Technology of China through the National Key Research and Development Program of China (Grant Nos. 2017YFA0303304 and 2016YFA0300601), the National Natural Science Foundation of China (Grant Nos. 92165208, 11874071, 91221202, and 91421303), and the Beijing Academy of Quantum Information Sciences (No. Y18G22).

**DATA AVAILABILITY**

The data that support the findings of this study are available within the article and its Supplementary Materials, and from the corresponding author upon reasonable request.

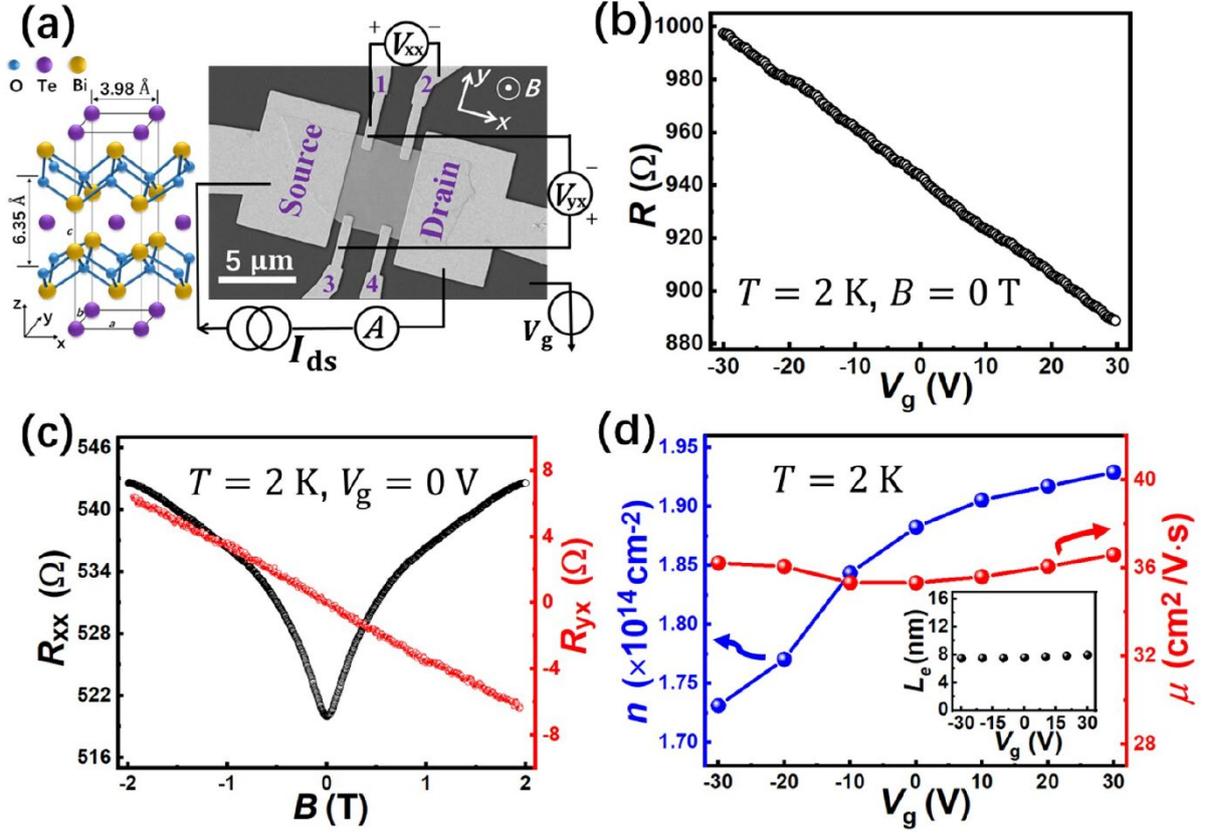

FIG. 1. (a) Left panel: Schematic diagram of the $Bi_2O_2Te$ crystal structure. Right panel: SEM image of the $Bi_2O_2Te$ Hall-bar device studied in this work and the measurement circuit setup. (b) Measured sheet resistance $R$ of the device vs. back-gate voltage $V_g$ at temperature $T = 2$ K and magnetic field $B = 0$ T. (c) Longitudinal resistance $R_{xx}$ and Hall resistance $R_{yx}$ of the device vs. $B$ at $T = 2$ K and $V_g = 0$ V. Note that in this work, $B$ is always applied perpendicular to the nanosheet as indicated in the right panel of Fig 1(a). (d) Measured sheet electron density $n$ and Hall mobility $\mu$ of the device vs. $V_g$ at $T = 2$ K. The inset shows the mean free path $L_e$ extracted from the measured values of $n$ and $\mu$.
12

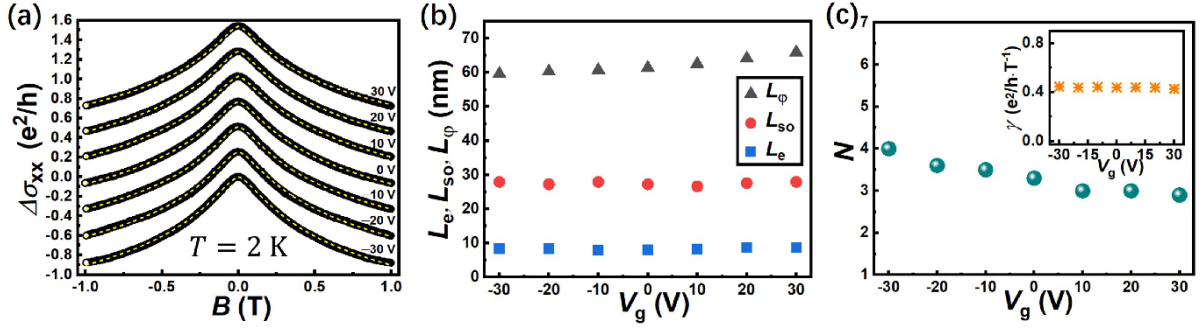

FIG. 2. (a) Magnetoconductivity $\Delta\sigma_{xx}$ (black opened circles) measured for the device at different $V_g$ and $T = 2$ K. The bottom black open circles show the data points measured at $V_g = -30$ V and the data points measured at all other values of $V_g$ are successively vertically offset for clarify. The yellow dashed lines represent the fits of the measurement data to Eq (1). (b) Dephasing length $L_\varphi$, spin relaxation length $L_{so}$ and mean free path $L_e$ in the nanosheet vs. $V_g$. (c) Effective conduction channel number $N$ in the nanosheet (main panel) and coefficient $\gamma$ in the correction term due to the LMR (inset) vs. $V_g$.

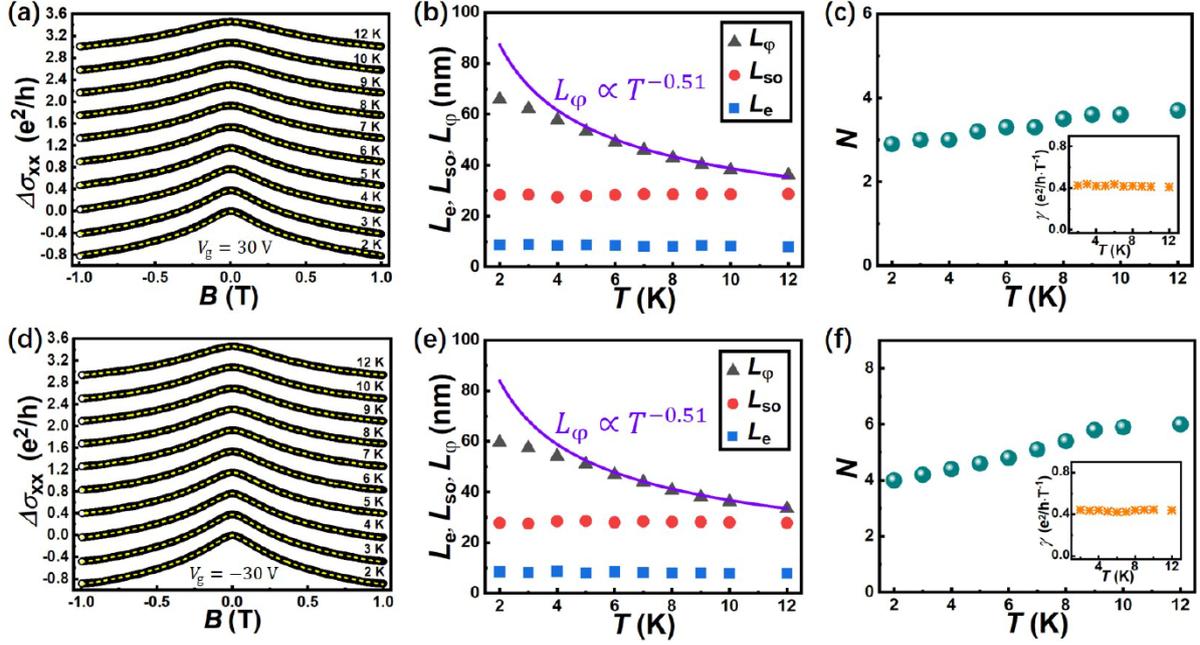

FIG. 3. (a) Magnetoconductivity $\Delta\sigma_{xx}$ (black opened circles) measured for the device at different temperatures $T$ and $V_g = 30$ V. The bottom black opened circles show the data points measured at $T = 2$ K and the data points measured at all other $T$ are successively vertically offset for clarify. The yellow dashed lines represent the fits of the measurement data to Eq (1). (b) Extracted $L_\varphi$, $L_{so}$ and $L_e$ vs. $T$ at $V_g = 30$ V. The solid line shows the power-law fit of $L_\varphi$ vs. $T$ by $L_\varphi \propto T^{-0.51}$. (c) Extracted $N$ (main panel) and $\gamma$ (inset) vs. $T$ at $V_g = 30$ V. (d)-(f) The same as (a)-(c) but for $V_g = -30$ V.

Supplementary Material for

# Electron transport properties of a narrow-bandgap semiconductor $Bi_2O_2Te$ nanosheet

Xiaobo Li[1,2], Haitian Su[1], and H. Q. Xu[1,3,*]

[1]*Beijing Key Laboratory of Quantum Devices, Key Laboratory for the Physics and Chemistry of Nanodevices, and School of Electronics, Peking University, Beijing 100871, China*

[2]*Academy for Advanced Interdisciplinary Studies, Peking University, Beijing 100871, China*

[3]*Beijing Academy of Quantum Information Sciences, Beijing 100193, China*

[*]Corresponding author; email: hqxu@pku.edu.cn

(Dated: June 1, 2022)

## CONTENTS

**Supplementary Note I:** Additional measurement data for the $Bi_2O_2Te$ nanosheet Hall-bar device studied in the main article

**Supplementary Note II:** Failure in analyzing the magnetotransport measurements shown in the main article using the quantum interference theory developed for a three-dimensional disordered system

**Supplementary Note III:** Fit of the measured magnetoconductivity data to the Hikami-Larkin-Nagaoka formula with the linear magnetoresistance (LMR) effect being taken into account in a larger field range of |B|<2 T.

**Supplementary Note IV:** Magnetotransport measurement data for additional $Bi_2O_2Te$ nanosheet Hall-bar devices

**Supplementary Note I: Additional measurement data for the $Bi_2O_2Te$ nanosheet Hall-bar device studied in the main article**

In this supplementary note, we will provide some additional measurement data for the $Bi_2O_2Te$ nanosheet Hall-bar device studied in the main article. Figure S1 shows an atomic force microscope (AFM) image of the device studied in the main article. The thickness ($d$~27 nm) of the $Bi_2O_2Te$ nanosheet employed in the device was determined by the step height in the height measurements along the white dashed line across an edge of the nanosheet.

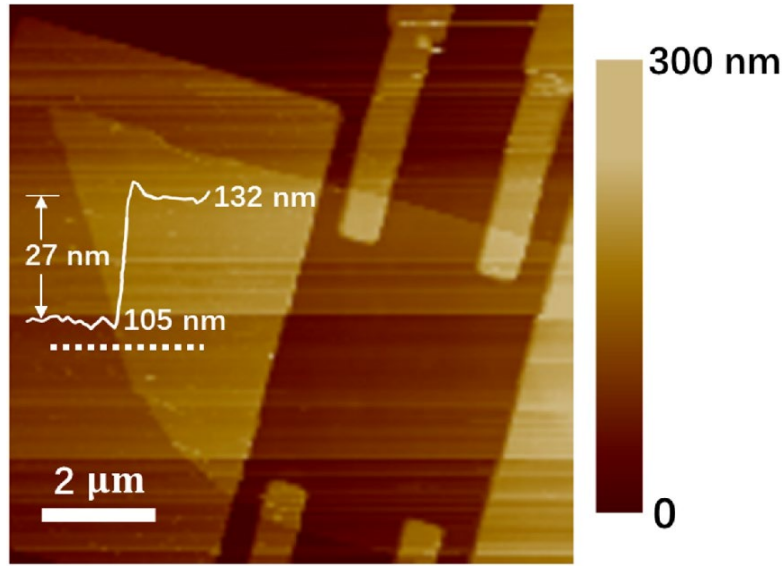

FIG. S1. Measured atomic force microscope (AFM) image of the $Bi_2O_2Te$ nanosheet device shown in Fig. 1(a) of the main article. The thickness of the nanosheet is determined to be $d$~27 nm by the height measurements along the white dashed line across an edge of the nanosheet.

Figure S2 shows the measured sheet resistance $R$ of the $Bi_2O_2Te$ nanosheet Hall-bar device studied in the main article as a function of temperature $T$ at zero back-gate voltage and zero magnetic field. It is seen that with decreasing $T$ from room temperature to about 200 K, the sheet resistance $R$ is gradually decreased, as a result of reduction in phonon scattering. At temperatures below about 200 K, the sheet resistance $R$ turns to increase with decreasing $T$. This resistance increase is dominantly due to reduction in thermally excited carriers at a lower temperature. At the very low temperatures, the change in the carrier density due to the contribution from thermal excitation becomes negligibly small and thus the sheet resistance $R$ is saturated.

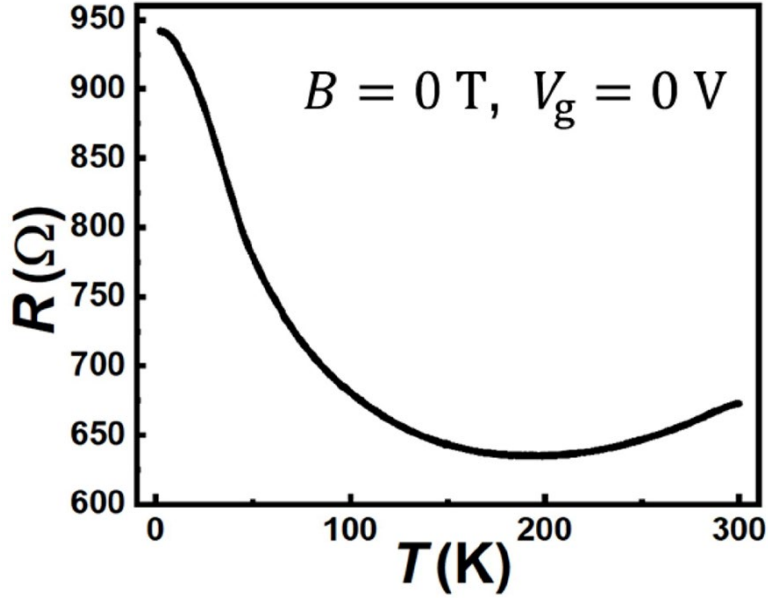

FIG. S2. Temperature $T$ dependence on the sheet resistance $R$ measured for the device shown in Fig. 1(a) of the main article at zero back-gate voltage and zero magnetic field.

**Supplementary Note II: Failure in analyzing the magnetotransport measurements shown in the main article using the quantum interference theory developed for a three-dimensional disordered system**

In this supplementary note, we will investigate whether the observed weak antilocalization (WAL) effect shown in the main article could also be analyzed by the quantum interference theory derived for a three dimensional (3D) disordered system and will confirm that the 3D quantum interference theory could not give a satisfactory description for the results of the magnetotransport measurements we have obtained and presented in the main article. For a mesoscopic 3D system, the magnetoresistivity is calculated by Fukuyama and Hoshino with the inelastic electron scattering, spin–orbit scattering and the Zeeman splitting of spin subbands being taken into account.[1] In case that the system is a superconductor but the environmental temperatures are above the superconducting transition temperature $T_c$, the contribution of fluctuational superconductivity to the magnetoresistivity should also be considered, because the fluctuational superconductivity gives a form of the magnetic field dependence of the resistance which is the same as the weak-localization contribution in the absence of spin–orbit scattering with a coefficient $\beta_{\text{Larkin}}(T)$ named as the Larkin e-e attraction strength.[2] Overall, the magnetoresistivity in a 3D system could be expressed as[1,2,3]

$$\frac{\Delta\rho(B)}{\rho^2(0)} = \frac{e^2}{2\pi^2\hbar}\sqrt{\frac{eB}{\hbar}}\left\{\frac{1}{2\sqrt{1-\gamma}}\left[f_3\left(\frac{B}{B_-}\right) - f_3\left(\frac{B}{B_+}\right)\right] - f_3\left(\frac{B}{B_2}\right) + \beta_{\text{Larkin}}(T)f_3\left(\frac{B}{B_\varphi}\right) - \sqrt{\frac{4B_{so}}{3B}}\left[\frac{1}{\sqrt{1-\gamma}}\left(\sqrt{t_+} - \sqrt{t_-}\right) + \sqrt{t} - \sqrt{t+1}\right]\right\}, \quad (1)$$

where $t = \frac{3B_\varphi}{2(2B_{so}-B_0)}$, $\gamma = \left[\frac{3g^*\mu_B B}{4eD(2B_{so}-B_0)}\right]^2$, $t_\pm = t + \frac{1}{2}(1 \pm \sqrt{1-\gamma})$, $B_\varphi = B_i + B_0$, $B_2 = B_i + \frac{1}{3}B_0 + \frac{4}{3}B_{so}$, $B_\pm = B_\varphi + \frac{1}{3}(2B_{so} - B_0)(1 \pm \sqrt{1-\gamma})$. Here, $g^*$ and $\mu_B$ are, respectively, the Landé g-factor and the Bohr magneton. $B_j$ is the characteristic field with $B_j = \frac{\hbar}{4eD\tau_j}$ [where j=$\varphi$, i, so, 0, and $\tau_j$, respectively, refer to the dephasing time, inelastic scattering time, spin relaxation time, and temperature-independent residual scattering time]. The characteristic time $\tau_j$ and characteristic length $L_j$ are connected by $L_j = \sqrt{D\tau_j}$, where $D$ is the electron diffusion coefficient. The function $f_3$ in Eq. (1) is an infinite series and could be approximately expressed by[4]

$$f_3(y) \approx 2\left[\sqrt{2+\frac{1}{y}} - \sqrt{\frac{1}{y}}\right] - \left[\left(\frac{1}{2}+\frac{1}{y}\right)^{-\frac{1}{2}} + \left(\frac{3}{2}+\frac{1}{y}\right)^{-\frac{1}{2}}\right] + \frac{1}{48}\left(2.03+\frac{1}{y}\right)^{-3/2}, \quad (2)$$

For an ordinary semiconductor, the superconducting fluctuation contribution to the magnetoresistivity could be neglected ($\beta_{\text{Larkin}} = 0$). In addition, when the transport measurements are performed in low magnetic fields, the Zeeman effect is small enough and could thus also be neglected ($\gamma \approx 0$), leaving $t_+ \approx t + 1$, $t_- \approx t$, $B_+ \approx B_\varphi + \frac{4}{3}B_{so} - \frac{2}{3}B_0$, and $B_- \approx B_\varphi$. As a result, the expression for the magnetoresistivity can be simplified to[5]

$$\frac{\Delta\rho(B)}{\rho^2(0)} = \frac{e^2}{2\pi^2\hbar}\sqrt{\frac{eB}{\hbar}}\left[\frac{1}{2}f_3\left(\frac{B}{B_\varphi}\right) - \frac{3}{2}f_3\left(\frac{B}{B_2}\right)\right], \quad (3)$$

where $B_\varphi = B_i + B_0$, $B_2 = B_i + \frac{1}{3}B_0 + \frac{4}{3}B_{so} = B_\varphi + \frac{4}{3}B_{so}^*$ with the effective spin-orbit field $B_{so}^* = B_{so} - \frac{1}{2}B_0$. The characteristic field $B_j$ and characteristic length $L_j^{(*)}$ are now connected by $B_j = \frac{\hbar}{4e\left[L_j^{(*)}\right]^2}$ (j=$\varphi$, i, so, 0).

For our semiconductor Bi$_2$O$_2$Te nanosheet Hall-bar device studied in the main article, the measured low-field magnetoresistivity could be analyzed by fitting the measurement data to Eq. (3) with dephasing length $L_\varphi$ and effective spin relaxation length $L_{so}^*$ as fitting parameters. Figure S3 shows such an analysis, where the black opened circles show the measured low-field magnetoresistivity data for the device shown in the main article at $T = 2$ K and $V_g = 0$ V. The red dashed line represents the best fit to the measurement data with Eq. (3). It is seen that the fit deviates strongly from the measurement data and, thus, the 3D quantum interference theory fails to analyze the magnetotransport measurements of our $Bi_2O_2Te$ nanosheet Hall-bar device studied in the main article.

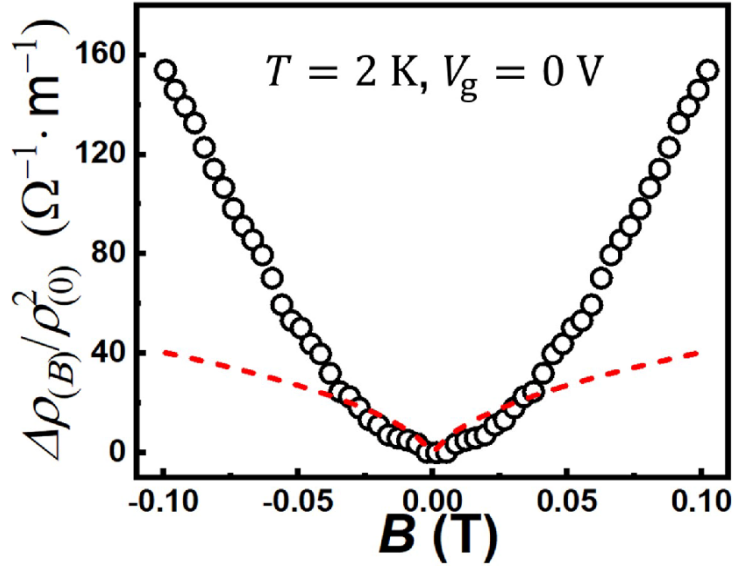

FIG. S3. Low-field normalized magnetoresistivity, $\Delta\rho(B)/\rho^2(0)$, (black open circles) measured for the $Bi_2O_2Te$ nanosheet Hall-bar device studied in the main article at $T = 2$ K and $V_g = 0$ V and theoretical fit (red dashed line) to Eq. (3). Here, $\Delta\rho(B) = \rho(B) - \rho(0)$, with $\rho$ as the bulk magnetoresistivity of the nanosheet. It is seen that the fit deviates strongly from the measurement data.

**Supplementary Note III: Fit of the measured magnetoconductivity data to the Hikami-Larkin-Nagaoka formula with the linear magnetoresistance (LMR) effect being taken into account in a larger field range of |B|<2 T.**

In the main article, the measured magnetoconductivity data have been analyzed using the Hikami-Larkin-Nagaoka (HLN) formula with the linear magnetoresistance (LMR) effect being taken into account[6,7] within $|B| < 1$ T. Considering that the LMR effect is prominently observed at higher magnetic fields, it is worthwhile to check whether the HLN formula with the LMR effect being taken into account could appropriately describe the measured magnetoconductivity data in a range of magnetic fields which includes the high magnetic field region in which the LMR effect appears clearly. Such a check is presented in Fig. S4, where the black opened circles show the measured magnetoresistivity data of the $Bi_2O_2Te$ nanosheet Hall-bar device studied in the main article, at $T = 2$ K and $V_g = 0$ V, within a larger magnetic field range ($|B| <$ 2 T). The green dashed line represents the fit result using the HLN formula with the LMR effect being taken into account. It is seen that the measured data points are excellently fitted by the theoretical formula within the whole magnetic field range. The fit yields dephasing length $L_\varphi \sim 61$ nm, spin relaxation length $L_{so} \sim 28$ nm, mean free path $L_e \sim 8$ nm, coefficient $\gamma \sim 0.44$ $e^2/h \cdot T^{-1}$ and the effective conduction channel number $N \sim 3.4$. All these values are, within a very small numerical uncertainty, the same as those extracted from the fit made within $|B| < 1$ T in the main article.

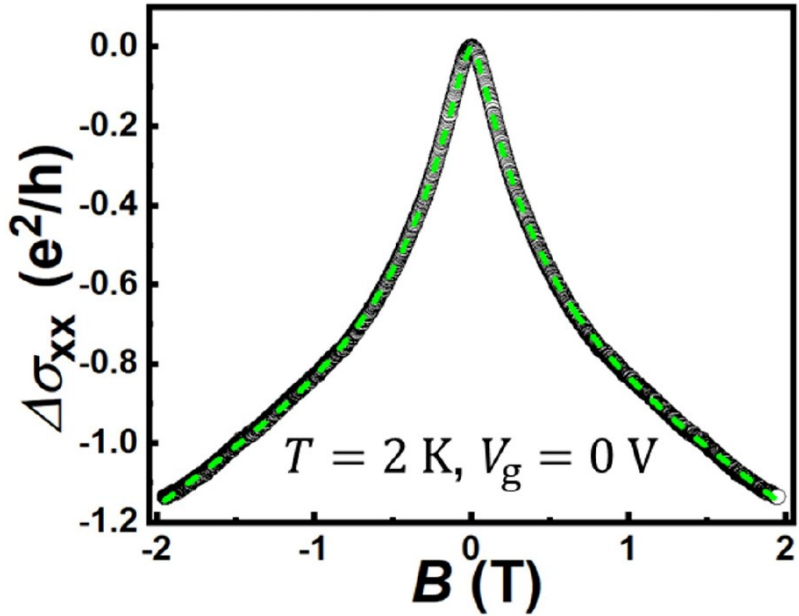

FIG. S4. Measured magnetoconductivity (black open circles) of the $Bi_2O_2Te$ nanosheet Hall-bar device studied in the main article at $T$ = 2 K and $V_g$ = 0 V and theoretical fit (green dashed line) to the HLN formula with the LMR effect being taken into account. Here, magnetic fields are taken into account in a larger range of $|B|$<2 T. It is seen that the measured data are still excellently fitted to the formula in $|B|$<2 T. The fit yields $L_\varphi \sim 61$ nm, $L_{so} \sim 28$ nm, $L_e \sim 8$ nm, $\gamma \sim 0.44$ $e^2/h \cdot T^{-1}$ and N~ 3.4. All the values are, within a very small numerical uncertainty, the same as those extracted from the fit made within $|B| < 1$ T in the main article.

**Supplementary Note IV: Magnetotransport measurement data for additional $Bi_2O_2Te$ nanosheet Hall-bar devices**

In this work, we have studied three $Bi_2O_2Te$ nanosheet Hall-bar devices. These devices are defined similarly in the device layout but made from the $Bi_2O_2Te$ nanosheets with different thicknesses. The device presented in the main article is made from a relatively thin nanosheet with a thickness of $d\sim27$ nm, the other two are made from relatively thick nanosheets with thicknesses $d\sim70$ nm and $d\sim110$ nm. All the magnetoresistivity measurements of the three devices at low temperatures show the weak antilocalization (WAL) characteristics and thus the presence of strong spin-orbit interaction in the nanosheets. Because in the devices made from the two thicker $Bi_2O_2Te$ nanosheets, the resistances and carrier densities can hardly be tuned with the back gates, we have in the main article presented our detailed measurements only for the device made from the thin $Bi_2O_2Te$ nanosheet (with $d\sim27$ nm) in which a clear gate tunability in carrier density and magnetoresistivity is observable. In this supplementary note, we show, as an example, an optical image of the device made from the nanosheet with ~70 nm in thickness, and the measured Hall and longitudinal resistance of the device as a function of perpendicularly applied magnetic field. Clearly, the measurements show the same magnetotransport characteristics as found for the device shown in the main article.

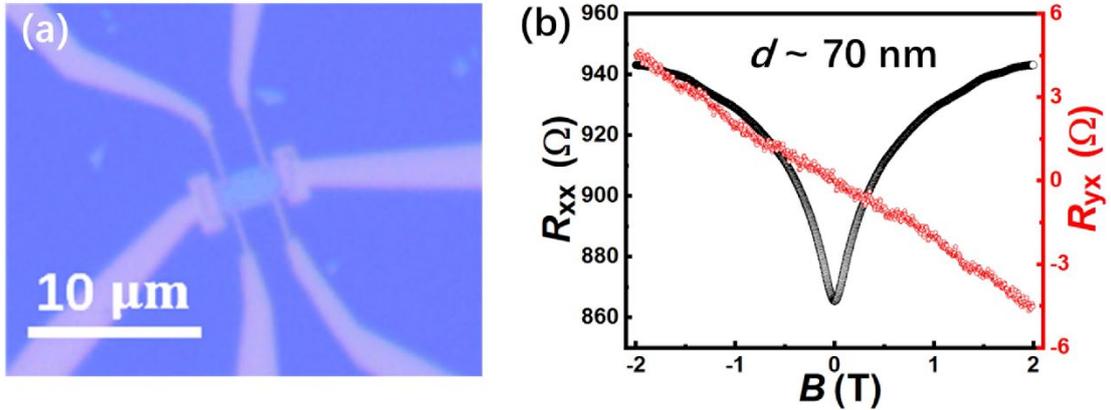

FIG. S5. (a) Optical image of a Hall-bar device made from a $Bi_2O_2Te$ nanosheet with a thickness of $d \sim 70$ nm. The width of the Hall bar and the distance between two longitudinal voltage probes in the device are $W$=1.3 μm and $L$=2.5 μm, respectively. (b) Measured longitudinal resistance $R_{xx}$ and Hall resistance $R_{yx}$ of the device vs. perpendicularly applied magnetic field $B$ at $T$ = 2 K and $V_g$ = 0 V. It is clearly seen that $R_{xx}$ displays a well-defined dip at low fields (a typical WAL characteristic), which confirms the presence of the strong spin-orbit interaction in the $Bi_2O_2Te$ nanosheet. The Hall resistance $R_{yx}$ decreases linearly with increasing $B$, confirming the transport carriers in the $Bi_2O_2Te$ nanosheet are of electrons. The extracted electron density and mobility from the measurements are $n \sim 2.7 \times 10^{14}$ cm$^{-2}$ and $\mu \sim 51$ cm$^2$/V·s.